# Breaking the Sensitivity Barrier in Fiber-Coupled Terahertz Spectroscopy

Xinghe Jiang[1, 2], Anika Tabassum[1, 2], Fanqi Shen[1, 2], Ali Jabbari[1, 2] & Mona Jarrahi[1, 2, *]

[1]Electrical and Computer Engineering Department, University of California; Los Angeles, 90095, USA.

[2]California NanoSystems Institute, University of California; Los Angeles, 90095, USA.

*Corresponding author. Email: mjarrahi@ucla.edu

**Abstract:** Fiber-coupled terahertz time-domain spectroscopy (THz-TDS) offers the flexibility and robustness needed for practical applications beyond controlled laboratory environments, but its sensitivity has remained substantially below that of free-space systems. Overcoming this limitation requires simultaneously achieving high-efficiency terahertz generation and detection, high-power femtosecond optical pulse delivery, and resilience to optical polarization fluctuations. Here, we demonstrate a fiber-coupled THz-TDS architecture that breaks this sensitivity barrier by integrating polarization-resilient plasmonic nanoantenna arrays with hollow-core antiresonant fibers. We establish a design principle that overcomes the intrinsic polarization sensitivity of plasmonic optoelectronic devices by exploiting the interplay between polarization-dependent optical absorption, ultrafast photocarrier transport, and local electric-field distributions. Through theoretical modeling and experimental validation, we show that engineering this interplay enables highly efficient terahertz generation and detection across different optical polarization states without compromising plasmonically enhanced optoelectronic efficiency. By suppressing optical dispersion and nonlinearities during high-power femtosecond pulse delivery, hollow-core antiresonant fibers further enable the superior optoelectronic properties of GaAs-based photoconductors to be fully exploited while preserving the flexibility and robustness of fiber coupling. The resulting system generates record-high terahertz radiation powers and achieves a 145-dB peak dynamic range in 60 seconds, representing a 300-fold improvement over state-of-the-art fiber-coupled THz-TDS systems. These advances overcome the longstanding trade-offs between plasmonic efficiency and polarization sensitivity, and between fiber-coupled operation and spectroscopic performance, establishing a new benchmark for high-sensitivity terahertz spectroscopy. This breakthrough opens new opportunities for probing weak light–matter interactions, characterizing strongly attenuating materials, and realizing high-throughput hyperspectral imaging and spatially distributed terahertz sensing beyond controlled laboratory environments.

## Introduction

Terahertz time-domain spectroscopy (THz-TDS) has emerged as a powerful technique for non-destructive material characterization[1-4], biomedical diagnostics[5-7], pharmaceutical and food quality control[8-11], industrial inspection and art conservation[12-14]. By generating ultrashort terahertz pulses and coherently detecting their electric fields, THz-TDS enables the non-contact characterization of materials by extracting their complex refractive indices over a broad spectral range. Among the different THz-TDS modalities, fiber-coupled THz-TDS systems have attracted significant interest because their fiber-based optical beam delivery makes them less susceptible to optical misalignment, mechanical vibrations, and environmental perturbations. This enhanced robustness, together with the flexibility of fiber-based optical delivery, makes fiber-coupled THz-TDS systems particularly attractive for portable and field-deployable applications outside controlled laboratory environments.

Erbium-doped mode-locked fiber lasers operating near 1550 nm have become particularly attractive for fiber-coupled THz-TDS systems because of their reliability and compatibility with mature telecommunications technology. In this wavelength range, the dispersion of silica fibers can be readily engineered, enabling all-fiber dispersion management, while the wide selection of commercially available Er-doped gain fibers and fiber-optic components facilitates the development of high-power femtosecond amplifiers. The 1550-nm band further benefits from a mature ecosystem of telecommunications components and established optical sampling systems, enabling rapid temporal scanning of terahertz pulses[15-20].

These attributes have motivated the extensive development of photoconductive terahertz sources and detectors compatible with 1550-nm excitation. Suitable photoconductive materials must simultaneously exhibit high carrier mobility for ultrafast photocarrier transport and high electrical resistivity to support large bias fields for efficient photoconductive terahertz generation and minimize Johnson–Nyquist noise for sensitive photoconductive terahertz detection. Particularly for terahertz detection, a short carrier lifetime is required to suppress the detrimental effects of slow photocarriers. InGaAs is the most widely used semiconductor for photoconductive terahertz antennas operating at 1550 nm because its bandgap energy enables efficient interband absorption at telecommunications wavelengths. However, InGaAs exhibits relatively low electrical resistivity, which negatively impacts terahertz generation efficiency and detection sensitivity. Furthermore, achieving a short carrier lifetime in InGaAs while maintaining high electrical resistivity and carrier mobility remains challenging.

Low-temperature epitaxial growth, which is highly effective in reducing the carrier lifetime of high-resistivity GaAs, makes InGaAs excessively n-type and conductive because of the high concentration of electrically active native defects[21,22]. To compensate for the background electrons, LT-InGaAs/InAlAs multilayer structures have been developed[23,24], in which the InAlAs layers provide additional carrier trapping and electrical isolation. This approach substantially increases electrical resistivity and enables a sub-picosecond response but requires precise compensation and suffers from reduced carrier mobility. Alternatively, ion implantation and heavy-ion irradiation can introduce recombination centers after material growth[25,26]. These techniques provide controllable defect densities but create randomly distributed lattice damage that increases carrier scattering and degrades mobility. More recently, transition-metal doping, particularly with Fe and Rh, has enabled sub-picosecond carrier lifetimes while preserving comparatively high mobility and sufficient electrical resistivity[27,28]. The incorporation of rare-earth elements, such as Er, has also been explored as an alternative approach for engineering carrier dynamics in InGaAs, as the introduced deep-level states can act as carrier trapping and recombination centers[29], albeit with considerable tradeoffs among electrical resistivity, carrier mobility, and carrier lifetime.

In comparison, semi-insulating (SI) GaAs and low-temperature-grown (LT) GaAs, which are commonly used for photoconductive terahertz sources and detectors operating at ~800 nm, offer significantly higher electrical resistivity than InGaAs-based photoconductors while maintaining high carrier mobility. LT-GaAs also provides sub-picosecond carrier lifetimes through comparatively straightforward low-temperature epitaxial growth followed by annealing[30]. In particular, the high dark resistance of SI-GaAs is a major advantage for photoconductive sources operating at ~800 nm, as it enables the application of much higher bias fields required for high-power terahertz generation without introducing large dark currents. Similarly, the high dark resistance of LT-GaAs is a major advantage for photoconductive detectors operating at ~800 nm, as it maintains low Johnson–Nyquist noise, which is essential for high-sensitivity terahertz detection.

Despite the favorable properties of GaAs-based photoconductive terahertz sources and detectors, the difficulty of delivering femtosecond pulses through optical fibers at ~800 nm has limited the development of high-performance fiber-coupled THz-TDS systems in this wavelength range. Conventional solid-core silica fibers exhibit normal material dispersion near 800 nm. Although specialized microstructured fibers can be engineered to provide anomalous dispersion, such designs generally require strong waveguide confinement and can consequently exhibit substantial optical nonlinearity. More fundamentally, the high peak powers of femtosecond pulses produce large nonlinear phase shifts through the Kerr effect in solid-core silica fibers. The resulting self-phase modulation, combined with chromatic dispersion and higher-order nonlinear effects, can substantially alter both the optical spectrum and the temporal pulse profile. Consequently, conventional solid-core fiber delivery of high-energy femtosecond pulses at ~800 nm is difficult to achieve without temporal broadening or pulse distortion, both of which directly degrade the efficiency and bandwidth of THz-TDS systems.

To address this bottleneck and utilize the superior performance of GaAs-based photoconductive antennas, we have developed a fiber-coupled THz-TDS system incorporating plasmonic nanoantenna arrays integrated with hollow-core antiresonant fibers (ARFs). Because light propagates predominantly through the air core of hollow-core ARFs, optical nonlinearities and chromatic dispersion are greatly reduced while propagation loss remains low[31-35]. This enables low-loss delivery of high-power femtosecond pulses at ~800 nm to photoconductive terahertz sources and detectors over long fiber lengths without significant pulse distortion or temporal broadening. Plasmonic structures significantly enhance the terahertz generation efficiency and detection sensitivity of photoconductive antennas by concentrating optical absorption near the contact electrodes, thereby enabling efficient collection of photogenerated carriers on sub-picosecond timescales[36-40]. However, a key challenge in integrating plasmonic photoconductive antennas with hollow-core ARFs is maintaining stable performance under optical polarization fluctuations. Such fluctuations can arise from mechanical motion and thermal variations in fiber delivery systems, as well as polarization instabilities in certain semiconductor and fiber lasers. Because plasmonic light–matter interactions are inherently polarization-dependent, these fluctuations can significantly affect the performance of conventional plasmonic photoconductive antennas, causing variations in radiated power exceeding two orders of magnitude[39-41] and compromising the stability and dynamic range of fiber-coupled THz-TDS systems.

Here, we demonstrate that plasmonic nanoantenna arrays can overcome intrinsic polarization sensitivity, enabling highly efficient and polarization-resilient terahertz generation and detection. We show, through analytical modeling and experimental characterization, that polarization-dependent variations in plasmonically enhanced optical absorption can be compensated by changes in the spatial overlap between photogenerated carriers and local electric fields. This interplay results in remarkably small variations in terahertz generation and detection sensitivity despite substantial differences in optical absorption distributions. By combining this polarization-resilient, plasmonics-enhanced response with high-power femtosecond pulse delivery through hollow-core ARFs, our approach enables high-stability, high-power, high-dynamic-range THz-TDS operation. We demonstrate record-high terahertz radiation power exceeding 2.1 mW and a peak dynamic range of 145 dB within a 60-s acquisition time, representing a 300-fold improvement over previously demonstrated fiber-coupled THz-TDS systems[42]. These advances overcome longstanding limitations of fiber-coupled THz-TDS systems, establishing a new benchmark for terahertz radiation power, dynamic range, and polarization resilience while paving the way for uncompromised spectroscopic performance beyond controlled laboratory environments.

# Results

The hollow-core ARF, with its air-guiding core and nearly circularly symmetric cladding structure, does not preserve a fixed optical polarization state, making the output polarization susceptible to mechanical and thermal perturbations. This presents a fundamental challenge for plasmonic devices, whose optical absorption is inherently polarization-dependent. To overcome this limitation, we leverage the polarization-dependent redistribution of plasmonically enhanced optical absorption relative to the local electric-field distribution, which compensates for variations in photocarrier generation efficiency and preserves the ultrafast photoconductive response across different optical polarization states.

Figure 1a illustrates the plasmonic nanoantenna arrays developed for polarization-resilient photoconductive terahertz generation on an SI-GaAs substrate. The grating-shaped plasmonic nanoantennas serve a dual function as broadband terahertz radiating elements and nanoscale optical field concentrators that enhance photocarrier generation in close proximity to the contact electrodes[43]. Under an applied bias voltage and femtosecond optical excitation, the photogenerated carriers drift toward the electrodes on a sub-picosecond timescale, producing a large transient photocurrent and strong pulsed terahertz radiation. The nanoantenna length is selected to support broadband terahertz radiation over the 0.1–10 THz frequency range, while their periodicity, width, and thickness are optimized to enhance optical field confinement near the electrodes. For optical pump polarization orthogonal to the gratings, the excitation of surface plasmon waves concentrates optical absorption along the electrode edges. In contrast, for polarization parallel to the gratings, the nanoantennas act as optical monopole antennas, concentrating the optical field and enhancing absorption near their tips. These distinct polarization-dependent optical absorption mechanisms provide a physical foundation for tailoring the nanoantenna geometry to achieve similar terahertz radiated fields across different optical polarization states, thereby enabling efficient and polarization-resilient terahertz generation. We establish this physical mechanism through analytical modeling, numerical simulations, and experimental characterization of the designed plasmonic nanoantenna terahertz source. The underlying physical principles and their quantitative agreement with experimental observations are detailed in the following analysis.

The far-field radiated terahertz field from each plasmonic nanoantenna is determined by its induced photocurrent. To quantify the impact of optical polarization on terahertz generation, we calculate the photocurrent induced on the plasmonic nanoantennas under both orthogonal and parallel optical polarizations using the Shockley–Ramo theorem. For a photogenerated carrier drifting in the substrate, the induced current on the plasmonic electrodes is $ev(r,t)E_w(r)$, where $e$ is the elementary charge, $v(r,t)$ is the drift velocity of the photocarrier, $r$ is the position vector, and $t$ is time. The weighting field $E_w(r)$ is determined by the electrode geometry and is obtained by setting the electrode of interest to unit potential while grounding all other electrodes. Integrating over the photo-generated electron and hole distributions, the total photocurrent $I_{ph}$ is given by:

$$I_{ph}(t) = e\int \left[p(r,t)v_p(r,t)\cdot E_w(r) - n(r,t)v_n(r,t)\cdot E_w(r)\right]d^3r \qquad (1)$$

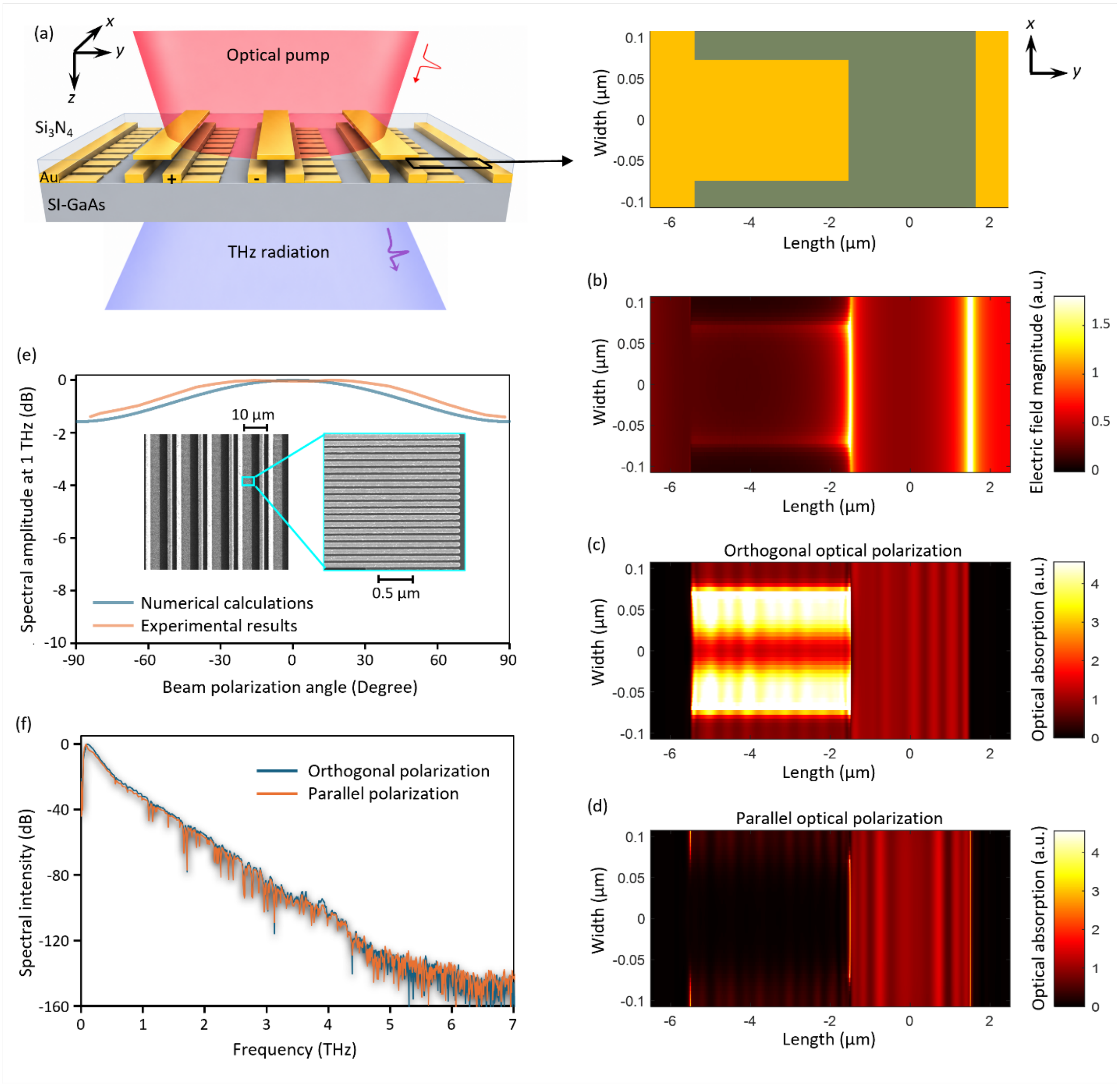

**Figure 1. Analysis of the polarization sensitivity of the plasmonic terahertz source.** (a) Schematic of the plasmonic terahertz source. (b) Magnitude of the bias electric field at a depth of 5 nm below the substrate surface for an applied bias voltage of 18 V, calculated using COMSOL. (c, d) Optical absorption profiles at a depth of 5 nm below the substrate surface for optical polarizations (c) orthogonal and (d) parallel to the plasmonic electrodes, calculated at a wavelength of 780 nm using Lumerical. (e) Numerically calculated and experimentally measured terahertz spectral amplitudes at 1 THz as a function of the optical polarization angle. Inset: Scanning electron microscopy image of the plasmonic terahertz source. (f) Resolved terahertz spectra for orthogonal and parallel optical pump polarizations.

where $p$ and $n$ are concentrations of photo-generated holes and electrons, with drift velocities $v_p$ and $v_n$, respectively. While the carrier concentrations are determined by optical absorption, their drift velocities are governed by the applied bias electric field, as detailed in the Methods section. Thus, obtaining a large photocurrent requires not only high optical absorption and a strong bias field but also a strong spatial overlap between the optical absorption and bias field distributions in the substrate. Figures 1b–d show the numerically calculated bias electric field and optical absorption distributions within the SI-GaAs substrate at a depth of 5 nm below the designed plasmonic nanoantenna arrays. For optical polarization orthogonal to the electrode orientation (Fig. 1c), the local optical absorption is enhanced along the electrode edges, whereas for parallel polarization (Fig. 1d), it is concentrated near the electrode tips. Although orthogonal polarization produces stronger optical absorption near the plasmonic contact electrodes, parallel polarization provides greater spatial overlap between the photoexcited carriers and the bias electric field. Consequently, the nanoantenna geometry is chosen to produce comparable transient photocurrents and terahertz radiation despite the differences in optical absorption distributions.

Using the bias electric field and optical absorption distributions, we numerically calculate the time-domain radiated terahertz field for different optical polarizations from the time-dependent photocurrent (see Methods section). The emitted terahertz spectrum is obtained by taking the time derivative of the photocurrent and applying a Fourier transform. Figure 1e shows the resulting spectral amplitude at 1 THz as a function of the optical polarization angle for an optical pump pulse with a full width at half maximum of 100 fs, predicting a 1.6 dB lower spectral amplitude for parallel polarization than for orthogonal polarization. To experimentally verify this prediction, we measured the THz-TDS spectra while varying the polarization angle at a fixed optical power incident on the plasmonic terahertz source. The measured amplitude at 1 THz is 1.4 dB below its maximum for parallel polarization, in close agreement with the numerical predictions. The small discrepancy between the numerical and experimental results is attributed to carrier screening, which is not accounted for in our theoretical model. At high photocarrier densities, the separation of electrons and holes under the bias field builds up an opposing electric field that reduces the net field in the absorption region and weakens the orthogonal-polarization response. Over the full measured bandwidth, the spectra for the two polarization states nearly overlap, as shown in Fig. 1f.

The same design methodology is extended to realize a polarization-resilient plasmonic terahertz detector. As illustrated in Fig. 2a, the detector incorporates plasmonic nanoantenna arrays fabricated on an LT-GaAs substrate, where the nanoantennas serve a dual function as broadband terahertz receiving elements that enhance the incident terahertz electric field within the photoconductive region and nanoscale optical field concentrators that enhance photocarrier generation near the contact electrodes[44]. Unlike the terahertz source, the detector operates without an externally applied bias voltage. Instead, the incident terahertz radiation induces a transient electric field across the photoconductive region, driving carriers generated by femtosecond optical probe pulses and producing a photocurrent proportional to the sampled terahertz electric field. The short carrier lifetime of LT-GaAs enables ultrafast temporal sampling of the incident terahertz waveform. The nanoantenna length and spacing are selected to enhance the induced terahertz electric field while maintaining broadband detection, whereas their periodicity, width, and thickness are tailored to enhance optical absorption near the electrodes. In particular, the separation between adjacent nanoantenna tips governs the local terahertz field enhancement and the distribution of photogenerated carriers within the high-field regions.

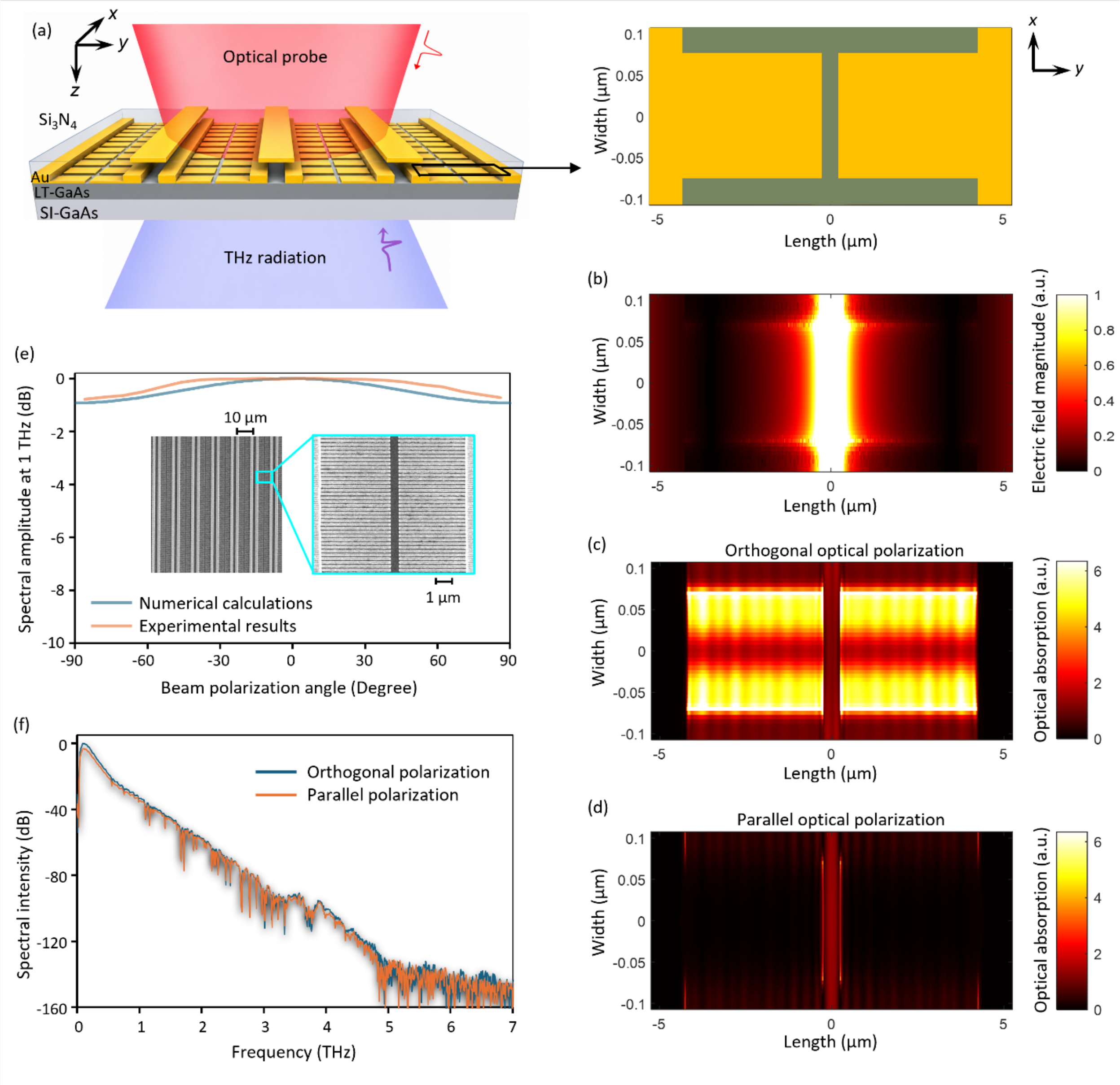

**Figure 2. Analysis of the polarization sensitivity of the plasmonic terahertz detector.** (a) Schematic of the plasmonic terahertz detector. (b) Magnitude of the induced terahertz electric field at a depth of 5 nm below the substrate surface for an incident terahertz field polarized parallel to the plasmonic electrodes, calculated using Lumerical. (c, d) Optical absorption profiles at a depth of 5 nm below the substrate surface for optical polarizations (c) orthogonal and (d) parallel to the plasmonic electrodes, calculated at a wavelength of 780 nm using Lumerical. (e) Numerically calculated and experimentally measured terahertz spectral amplitudes at 1 THz as a function of the optical polarization angle. Inset: Scanning electron microscopy image of the plasmonic terahertz detector. (f) Resolved terahertz spectra for orthogonal and parallel optical probe polarizations.

Figures 2b–d show the numerically calculated terahertz electric field and optical absorption distributions within the LT-GaAs substrate at a depth of 5 nm below the designed plasmonic nanoantenna arrays. Similar to the terahertz source, the plasmonic detector exhibits distinct optical absorption mechanisms depending on the probe polarization state. For optical polarization orthogonal to the nanoantenna gratings, surface plasmon excitation concentrates optical absorption along the electrode edges (Fig. 2c). In contrast, for parallel polarization, the nanoantennas act as optical dipole antennas, concentrating the optical field and enhancing absorption near their tips (Fig. 2d). Although these two polarization states produce substantially different spatial distributions of photogenerated carriers, their contributions to the detected photocurrent are determined by the spatial overlap between the photocarrier distributions and the terahertz electric field induced within the photoconductive region. The nanoantenna geometry is therefore engineered to account for polarization-dependent variations in optical absorption through corresponding changes in this spatial overlap, enabling comparable photocurrent responses for orthogonal and parallel optical probe polarizations.

Using the induced terahertz electric field and optical absorption distributions, we numerically calculate the detected photocurrent as a function of the time delay between the optical pump and probe pulses, which represents the detected time-domain terahertz electric field (see Methods section). The detected terahertz spectrum is subsequently obtained by taking the Fourier transform of the time-domain terahertz electric field. Figure 2e shows the resulting spectral amplitude at 1 THz as a function of the optical polarization angle for an incident terahertz pulse with a full width at half maximum of 0.6 ps, predicting a 0.9 dB lower spectral amplitude for parallel polarization than for orthogonal polarization. To experimentally verify this prediction, we measured THz-TDS spectra while varying the polarization angle at a fixed optical power incident on the plasmonic terahertz detector. The measured amplitude at 1 THz is 0.8 dB below its maximum for parallel polarization, in close agreement with the numerical predictions. The small discrepancy between the numerical and experimental results is attributed to carrier screening, an effect not included in our theoretical model. Nevertheless, across the full measured bandwidth, the spectra obtained for the two polarization states remain nearly identical, as shown in Fig. 2f. Together, these results demonstrate that the engineered plasmonic nanoantenna arrays enable polarization-resilient terahertz generation and detection, mitigating the impact of polarization fluctuations introduced by the hollow-core ARFs and supporting stable THz-TDS operation.

The plasmonic nanoantenna array terahertz source and detector were fabricated on SI-GaAs and LT-GaAs substrates, respectively (see Methods section). A custom package was developed to provide robust optical coupling between the hollow-core ARF and the plasmonic nanoantenna arrays (Fig. 3a, inset), enabling a modular and flexible terahertz scan head. Figure 3a shows the terahertz power radiated by the plasmonic nanoantenna array source at different optical power levels and bias voltages, measured using a calibrated pyroelectric detector (PTB 19, THz 30). Record-high terahertz power levels exceeding 2.1 mW were achieved, surpassing those previously demonstrated by fiber-coupled pulsed terahertz sources despite operating at an order of magnitude lower bias voltage. Using the fiber-coupled plasmonic nanoantenna array terahertz source and detector, we constructed the THz-TDS system shown in Fig. 3b. The system consists of a fiber-coupled optical unit in which the output of a 780-nm femtosecond laser is split into pump and probe beams, which pass through a controllable optical delay and are subsequently coupled into separate hollow-core ARFs that deliver the optical pulses to the plasmonic nanoantenna array terahertz source and detector, respectively.

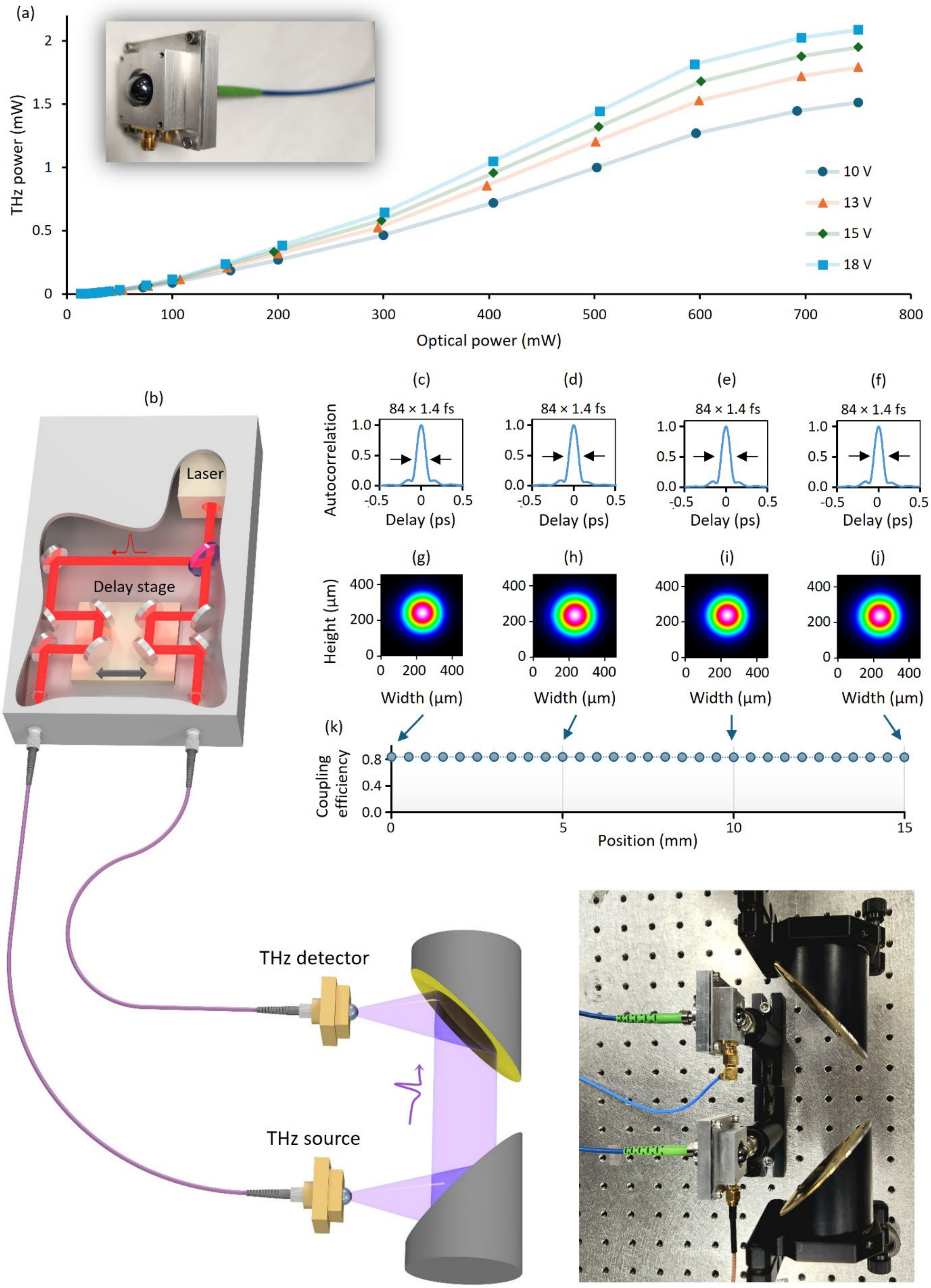
(a)
THz power (mW)
Optical power (mW)
10 V
13 V
15 V
18 V
(b)
Laser
Delay stage
THz detector
THz source
(c)
(d)
(e)
(f)
84 × 1.4 fs
Autocorrelation
Delay (ps)
(g)
(h)
(i)
(j)
Height (μm)
Width (μm)
(k)
Coupling efficiency
Position (mm)

**Figure 3.** (a) Radiated terahertz power of the hollow-core ARF-coupled plasmonic nanoantenna arrays as a function of the optical power and bias voltage. Inset shows the photograph of a hollow-core ARF-coupled plasmonic nanoantenna array. (b) Experimental setup of the hollow-core ARF-coupled THz-TDS system. (c-f) The autocorrelation traces and (g-j) the output beam profiles measured at delay-line positions of 0, 5, 10, and 15 mm. The autocorrelation traces maintained a full width at half maximum of 119 fs, corresponding to an optical pulse width of 84 fs, assuming a Gaussian temporal profile. With the beam profiler positioned 20 mm from the fiber tip, the measured beam profiles exhibited a consistent single-mode spatial profile. These results demonstrate that the hollow-core ARF maintains stable optical coupling, spatial beam quality, and femtosecond pulse duration throughout delay-line scanning, supporting robust THz-TDS operation in the subsequent experiments. (k) Optical coupling efficiency to the hollow-core ARF at different delay-line positions.

Unlike conventional optical fibers with solid silica cores, hollow-core ARFs use highly reflective microstructures to confine light within an air core. Consequently, light propagating through a hollow-core ARF experiences negligible dispersion and optical nonlinearity, enabling the use of long fibers without significantly affecting the optical pulse width. In our THz-TDS system, hollow-core ARFs with lengths of 1.8 m and 1.4 m were used for the optical pump and probe paths, respectively. No noticeable changes in pulse width, pulse shape, or beam profile were observed, even when translating the optical delay line over a 15-mm range, corresponding to a 200-ps temporal scan (Fig. 3c–j). Furthermore, a stable optical coupling efficiency of 83% was maintained throughout the optical scan range. This high coupling efficiency is enabled by the relatively large fiber core diameter and reduced back-reflection associated with coupling light into an air core rather than a solid core (Fig. 3k).

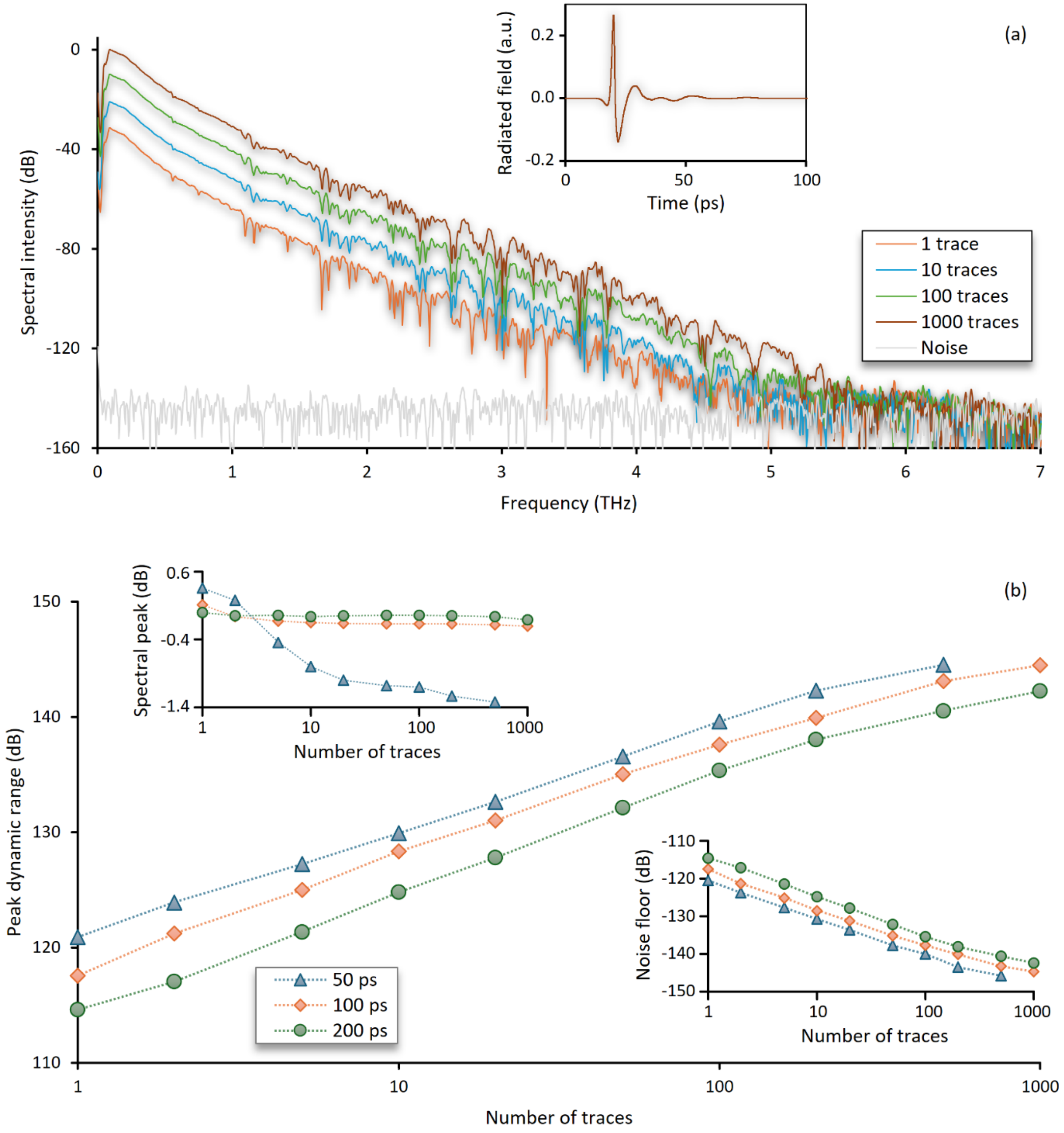

**Figure 4.** (a) Measured terahertz power spectra as a function of the number of captured traces for a 100-ps scan range. Inset shows the radiated terahertz electric field over a 100-ps scan range using 1000 captured traces. The noise floors were measured by blocking the pump beam coupled to the terahertz source. The terahertz spectra are shifted to align the noise floor at the same level. (b) Peak dynamic range as a function of the number of captured traces with scan ranges of 50, 100, 200 ps. Insets: spectral intensity peak and noise floor as a function of the number of captured traces with scan ranges of 50, 100, 200 ps.

Figure 4a shows the terahertz spectra obtained using the fiber-coupled THz-TDS system as a function of the number of acquired traces. The optical delay range was adjusted to provide a temporal scan range of 100 ps (Fig. 4a, inset). The optical pump and probe powers driving the terahertz source and detector were set to 600 and 150 mW, respectively, to maximize the system dynamic range within a total optical power budget of 900 mW (see Supplementary Fig. S2). The peak dynamic range increases from 118 dB to 145 dB as the number of acquired traces increases from 1 to 1,000, corresponding to acquisition times ranging from 0.19 to 190 s.

| THz-TDS systems | Bias voltage (V) | Radiated THz power (mW) | Peak dynamic range (dB) | Acquisition time (s) | Scan range (ps) |
|---|---|---|---|---|---|
| This work | 18 | 2.1 | 132 | 1.5 | 50 |
| Menlo Tera K15[46] | 200 | 0.25 | 110 | 1.5 | 50 |
| This work | 18 | 2.1 | 145 | 60 | 50 |
| Toptica TeraFlash pro[47] | 200 | 0.4 | 120 | 60 | 50 |

**Table 1.** Performance comparison of the THz-TDS system developed in this work with state-of-the-art THz-TDS systems.

This increase in dynamic range is 3 dB below the theoretical prediction. Under ideal conditions, the time-domain terahertz signal remains unchanged across all acquired traces, preserving the deterministic signal amplitude upon averaging. In contrast, the noise power, dominated by Johnson–Nyquist noise in the terahertz detector[45], is expected to decrease by 10 dB per decade with increasing numbers of averaged traces, owing to the uncorrelated nature of the noise across individual traces. As shown in Fig. 4b, both the signal and noise power levels deviate from their theoretical predictions, resulting in a sublinear increase in dynamic range with the number of averaged traces. We attribute these deviations to imperfect repeatability of the optical delay line, which introduces temporal misalignment between acquired traces, reducing the averaged terahertz signal amplitude and introducing additional noise that limits the reduction in noise power through averaging. This hypothesis is further supported by the dependence of these deviations on the temporal scan range, with larger deviations observed for a 50-ps scan range and smaller deviations for a 200-ps scan range compared with the 100-ps scan range as illustrated in Fig. 4b (see Supplementary Fig. S1). This trend is consistent with reduced repeatability of the optical delay line over shorter scan ranges, where a larger fraction of the scan occurs during the acceleration and deceleration phases of the delay stage. Although higher-accuracy optical delay stages could further improve the achievable dynamic range, the demonstrated fiber-coupled THz-TDS system offers more than two orders of magnitude higher dynamic range than state-of-the-art systems, even with these limitations, as illustrated in Table 1[42,46,47]. This significant enhancement is enabled by the high terahertz generation and detection efficiencies of the polarization-resilient plasmonic nanoantenna arrays, combined with the high optical power-handling capability of hollow-core ARFs.

## Discussion

Our findings establish a new approach to engineering polarization-resilient plasmonic optoelectronic devices by exploiting the dependencies between nanoscale optical field confinement and ultrafast photocarrier transport. Although plasmonic light–matter interactions are inherently polarization-dependent, our results demonstrate that this dependence does not necessarily translate into corresponding variations in the macroscopic optoelectronic response. Specifically, distinct optical absorption distributions induced by orthogonal polarization states can produce comparable transient photocurrents when the spatial overlap between photogenerated carriers and local electric fields is appropriately engineered. This finding extends the conventional paradigm of plasmonic device optimization beyond maximizing optical absorption or electromagnetic field enhancement, establishing the spatial distribution of photocarrier generation relative to the local electric field as a critical design parameter. By exploiting this additional degree of freedom, polarization resilience can be achieved while preserving the strong light–matter interactions that underpin plasmonically enhanced optoelectronic efficiency.

Another important advance enabled by this work is overcoming a longstanding trade-off between the optoelectronic properties of photoconductive materials and the advantages of fiber-based optical delivery. Conventional fiber-coupled THz-TDS systems have predominantly operated at telecommunications wavelengths, where mature fiber-optic technologies enable efficient delivery of femtosecond optical pulses but constrain the choice of photoconductive materials. In particular, InGaAs-based photoconductors operating near 1550 nm face competing requirements for high carrier mobility, high electrical resistivity, and ultrafast carrier recombination, limiting terahertz generation efficiency and detection sensitivity. In contrast, GaAs-based photoconductors operating near 800 nm offer a more favorable combination of these properties, but dispersion and optical nonlinearities in conventional solid-core fibers have restricted their use in high-power fiber-coupled systems. By integrating hollow-core antiresonant fibers with plasmonic nanoantenna arrays, our approach overcomes this limitation, enabling high-power femtosecond excitation of GaAs-based plasmonic photoconductive devices without significant optical pulse distortion or temporal broadening. This architecture decouples the selection of photoconductive materials from the constraints imposed by conventional optical fibers, allowing the excitation wavelength to be chosen primarily based on the intrinsic optoelectronic properties of the active material[48–50]. This capability could have broader implications beyond terahertz spectroscopy, particularly for ultrafast optoelectronic systems requiring high-power femtosecond excitation at wavelengths that are challenging for conventional fiber-based optical delivery.

The combination of polarization-resilient plasmonic nanoantenna arrays and hollow-core fiber delivery also challenges the conventional trade-off between spectroscopic performance and system robustness in terahertz time-domain spectroscopy. Historically, high-dynamic-range THz-TDS systems have relied on free-space optical architectures, whereas fiber-coupled systems have offered greater flexibility and reduced alignment complexity at the expense of terahertz generation efficiency and detection sensitivity. Our results demonstrate that these limitations are not intrinsic to fiber-coupled operation. By combining efficient photocarrier generation and collection in GaAs-based plasmonic devices with high-power optical delivery through hollow-core fibers, the developed system achieves a record-high radiated terahertz power of 2.1 mW and a peak dynamic range of 145 dB, exceeding the performance of state-of-the-art fiber-coupled THz-TDS systems by more than two orders of magnitude. Importantly, the advantages of the demonstrated fiber-coupled THz-TDS architecture are not limited to the specific plasmonic nanoantenna arrays developed in this work. The radiated terahertz power and THz-TDS dynamic range could be further enhanced by incorporating 3D plasmonic contact electrodes[51,52] and plasmonic nanocavities[53–55] into the active regions of the terahertz source and detector, as well as by employing plasmonic contact electrodes with lower energy barriers[56], while preserving polarization-resilient operation. Furthermore, although the demonstrated source achieves record-high radiated terahertz power among fiber-coupled THz-TDS systems at an order-of-magnitude lower bias voltage than previously reported sources, bias-free terahertz sources based on plasmonic nanoantenna arrays could also be integrated into this architecture while maintaining a high dynamic range[57–59].

The substantially enhanced dynamic range is particularly important for spectroscopy of strongly attenuating materials and detection of weak spectral features, where limited dynamic range can restrict the accessible frequency bandwidth and compromise the accuracy of extracted material properties. Beyond single-channel spectroscopy, the demonstrated architecture offers opportunities to extend high-performance terahertz measurements to parallel and spatially distributed sensing and imaging systems. The ability to deliver high-power femtosecond pulses through flexible optical fibers is particularly attractive for plasmonic terahertz focal-plane arrays[60,61] and source arrays[62], where sufficient optical excitation power is essential for maintaining high sensitivity across multiple spatial channels. Integrating hollow-core fiber delivery with polarization-resilient plasmonic nanoantenna arrays could enable high-throughput hyperspectral terahertz imaging without sequential mechanical scanning of individual spatial pixels, opening new possibilities for high-sensitivity, scalable, and robust terahertz sensing and imaging.

# Methods

### Fabrication of plasmonic nanoantenna arrays

The terahertz source, comprising plasmonic nanoantenna arrays (Fig. 1a) with a total active area of 1 mm × 1 mm, was fabricated on an SI-GaAs substrate. The terahertz detector, comprising plasmonic nanoantenna arrays (Fig. 2a) with a total active area of 250 µm × 250 µm, was fabricated on an LT-GaAs substrate. Both devices were fabricated using the same process. First, plasmonic nanoantennas with a width of 140 nm and a periodicity of 210 nm were patterned by electron-beam lithography, followed by the deposition of 3/47 nm Ti/Au and lift-off. Next, the bias lines and contact pads were defined by photolithography, followed by the deposition of 20/180 nm Ti/Au and lift-off. A $Si_3N_4$ antireflection coating was subsequently deposited by plasma-enhanced chemical vapor deposition. Shadow metals were then patterned on top of the antireflection coating using photolithography, followed by metal deposition and lift-off. Finally, vias were opened through the dielectric layers to expose the contact pads for wire bonding of the fabricated terahertz devices.

**Numerical Calculation of the Photocurrent Induced in Plasmonic Nanoantenna Arrays**

According to the Shockley–Ramo theorem, the photocurrent induced at the plasmonic electrodes of the terahertz source is given by:

$$I_{ph}(t) = e\int\left[p(r,t)v_p(r,t)\cdot E_w(r) - n(r,t)v_n(r,t)\cdot E_w(r)\right]d^3r \quad (1)$$

where the first and second terms represent the contributions from holes and electrons, with carrier concentrations $p$ and $n$, and drift velocities $v_p$ and $v_n$, respectively. $E_w$ is the weighting field determined by the geometry of the nanoantenna arrays (calculated numerically by COMSOL under unity bias voltage). The drift velocities are obtained from the Drude model:

$$m_p^*\frac{\partial v_p(r,t)}{\partial t} + m_p^*\frac{v_p(r,t)}{\tau_{sc,p}} = eE_{bias}(r) \quad (2)$$

$$m_n^*\frac{\partial v_n(r,t)}{\partial t} + m_n^*\frac{v_n(r,t)}{\tau_{sc,n}} = -eE_{bias}(r) \quad (3)$$

where $E_{bias}(r)$ is the bias electric field (calculated numerically by COMSOL under an 18-v bias voltage). $\tau_{sc,p}$ and $\tau_{sc,n}$ are the momentum scattering time of holes and electrons respectively, which are functions of the carrier mobility $\mu$ and the effective mass $m^*$: $\tau_{sc,p} = \frac{m_p^*\mu_p}{e}$ and $\tau_{sc,n} = \frac{m_n^*\mu_n}{e}$. For a bias field that is constant in time, the photocarrier drift velocity is calculated as:

$$v_p(r,t) = \mu_p E_{bias}(r)\left[1 - \exp\left(-\frac{t}{\tau_{sc,p}}\right)\right] \quad (4)$$

$$v_n(r,t) = \mu_n E_{bias}(r)\left[1 - \exp\left(-\frac{t}{\tau_{sc,n}}\right)\right] \quad (5)$$

The photocarrier concentrations $n$ and $p$, are obtained by solving the charge-continuity equation:

$$\frac{\partial n(r,t)}{\partial t} = -\nabla \cdot [n(r,t)v_n(r,t)] + G(r,t) - R_n(r,t) \quad (6)$$

$$\frac{\partial p(r,t)}{\partial t} = -\nabla \cdot \left[p(r,t)v_p(r,t)\right] + G(r,t) - R_p(r,t) \quad (7)$$

where $G(t)$ and $R(t)$ denote the photocarrier generation and recombination rates, respectively. For electron and hole lifetimes $\tau_{rec,n}$ and $\tau_{rec,p}$, the recombination terms are $R_n(r,t) = \frac{n(r,t)}{\tau_{rec,n}}$ and $R_p(r,t) = \frac{p(r,t)}{\tau_{rec,p}}$. The generation rate $G(t)$ is determined by the optical excitation. Assuming a Gaussian optical pulse, the photocarrier generation rate is:

$$G(r,t) = \frac{\alpha F_{pump}(r)}{h\nu}\frac{1}{\sqrt{2\pi}T}\exp\left[-\frac{\left(t-\tau_{opt}\right)^2}{2T^2}\right] \quad (8)$$

where $\alpha$ is the optical absorption coefficient of the substrate and $h\nu$ is the photon energy. The values $T = 42.47\ fs$ and $\tau_{opt} = 500\ fs$ yield a pulse with full-width half maximum of 100 fs centered at 500 fs on simulation time axis. $F_{pump}(r)$ is the local optical fluence, which is numerically solved by Lumerical at 780 nm. We use a split-step method to numerically solve the charge-continuity equation. Specifically, for each time increment $\Delta t$, the changes in carrier density due to generation and recombination are calculated first, followed by carrier transport to obtain the time-dependent electron and hole concentrations $n(r,t)$ and $p(r,t)$.

The time-dependent photocurrent is calculated using the Shockley–Ramo theorem (Eq. 1), and the far-field radiated terahertz electric field is obtained by taking the time derivative of the photocurrent. The radiated terahertz spectrum is then calculated by applying a Fourier transform to the resulting terahertz field, and its amplitude at 1 THz is used for comparison with the experimental results.

As with the plasmonic terahertz source, the photocurrent induced in the plasmonic terahertz detector is calculated using the Shockley–Ramo theorem. However, in the absence of an external bias voltage, the electric field induced by the incident terahertz radiation, $E_{THz}(r,t)$, drives the photogenerated carriers within the active region of the device. The induced terahertz electric field distribution is numerically calculated using Lumerical for an incident terahertz pulse with a Gaussian envelope and a full width at half maximum (FWHM) of 0.6 ps. The calculated field distribution is then used to determine the photocarrier drift velocities using the Drude model:

$$m_p^* \frac{\partial v_p(r,t)}{\partial t} + m_p^* \frac{v_p(r,t)}{\tau_{sc,p}} = eE_{THz}(r,t) \tag{9}$$

$$m_n^* \frac{\partial v_n(r,t)}{\partial t} + m_n^* \frac{v_n(r,t)}{\tau_{sc,n}} = -eE_{THz}(r,t) \tag{10}$$

The timing of the optical pulse, $\tau_{opt}$, corresponds to the sampling delay of the terahertz pulse. For each sampling delay, the time-dependent photocarrier density and photocurrent are calculated using the split-step method, and the resulting photocurrent is integrated to estimate the signal measured by the THz-TDS data acquisition system. The calculated photocurrent as a function of sampling delay, which represents the detected time-domain terahertz electric field, is then Fourier-transformed to obtain the terahertz spectrum.

**Hollow-core ARF-coupled THz-TDS system**

The optical output of a 780-nm femtosecond laser (Toptica FemtoFiber Ultra 780), delivering 84-fs pulses at a repetition rate of 100 MHz, is split into pump and probe branches. Both beams are directed through an optical delay line (Newport DL 125) to double the scan speed. The delayed pump and probe beams are then coupled into 1.8-m and 1.4-m hollow-core ARFs connected to the plasmonic terahertz source and detector, respectively. A pair of off-axis gold-coated parabolic mirrors, each with a focal length of 2 inches, is used to collimate the generated terahertz radiation and focus it onto the terahertz detector. The 18-V bias applied to the plasmonic terahertz source is modulated to reduce flicker noise. The resulting photocurrent from the terahertz detector is amplified by a transimpedance amplifier (FEMTO DLPCA-200) and subsequently measured using a lock-in amplifier (Zurich Instruments MFLI). The time-domain terahertz electric field is obtained by recording the lock-in-detected photocurrent as a function of the optical delay.

**Acknowledgements**

The authors acknowledge funding from the Moore Foundation and AFOSR Award #FA9550-25-1-0358. Development of the plasmonic nanoantenna arrays used in this work was supported by the Department of Energy (grant # DE-SC0016925).

**Contributions**

X.J., F.S. and A.J. designed the polarization resilient plasmonic nanoantenna arrays and performed the numerical studies; A.T. fabricated and packaged the plasmonic nanoantenna arrays; X.J. designed the hollow-core ARF-coupled plasmonic nanoantenna array package and built the THz-TDS setup, performed the experiments and data analysis; M.J. initiated and supervised the research. All authors contributed to the preparation of the manuscript.

# Supplementary Information

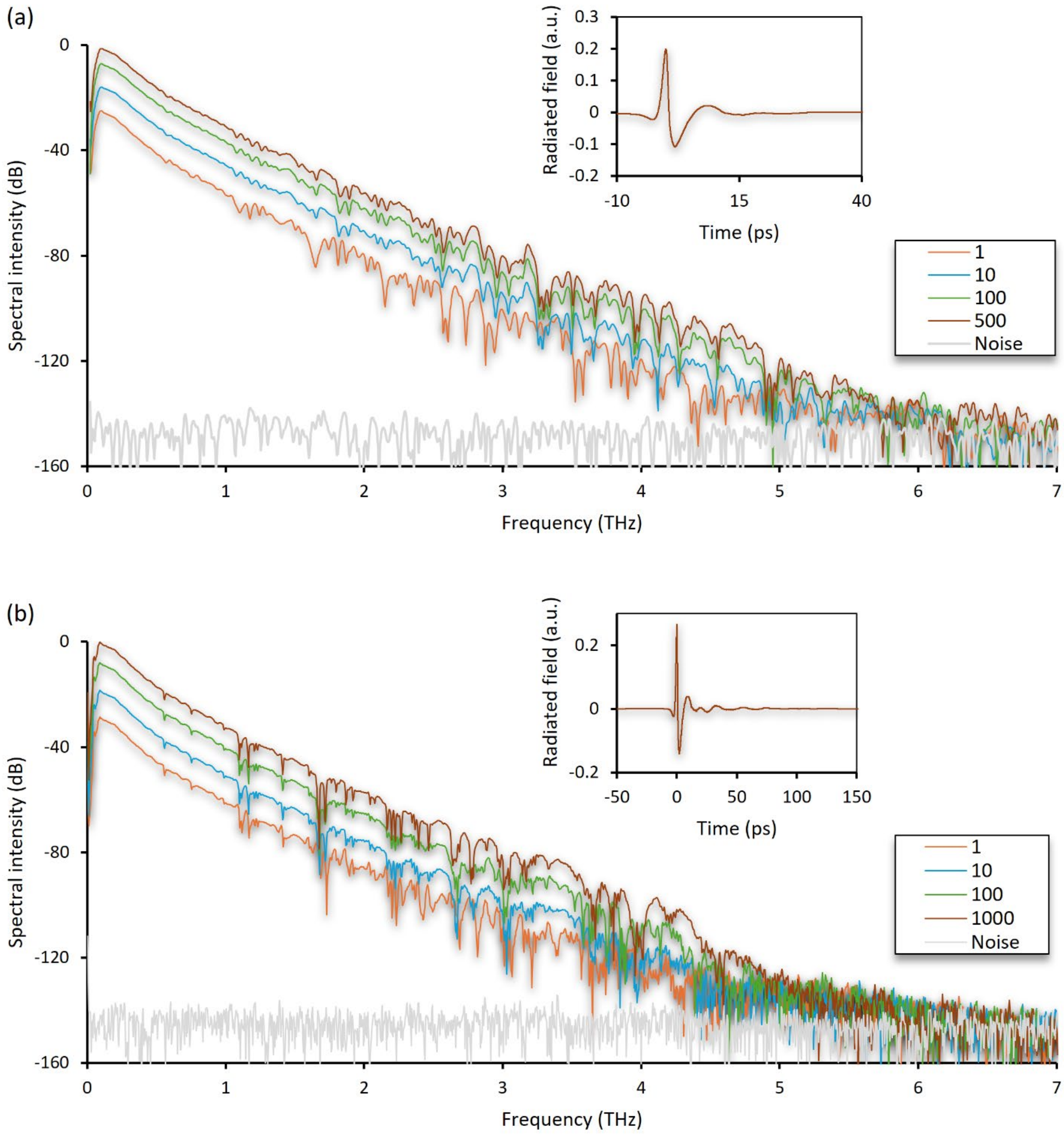

**Figure S1.** Terahertz power spectra as a function of the number of acquired traces for scan ranges of (a) 50 ps and (b) 200 ps. The insets show the corresponding time-domain terahertz electric fields obtained from 1,000 acquired traces. The noise floors were measured by blocking the optical pump beam coupled to the terahertz source. The spectra were vertically shifted to align their noise floors at the same level.

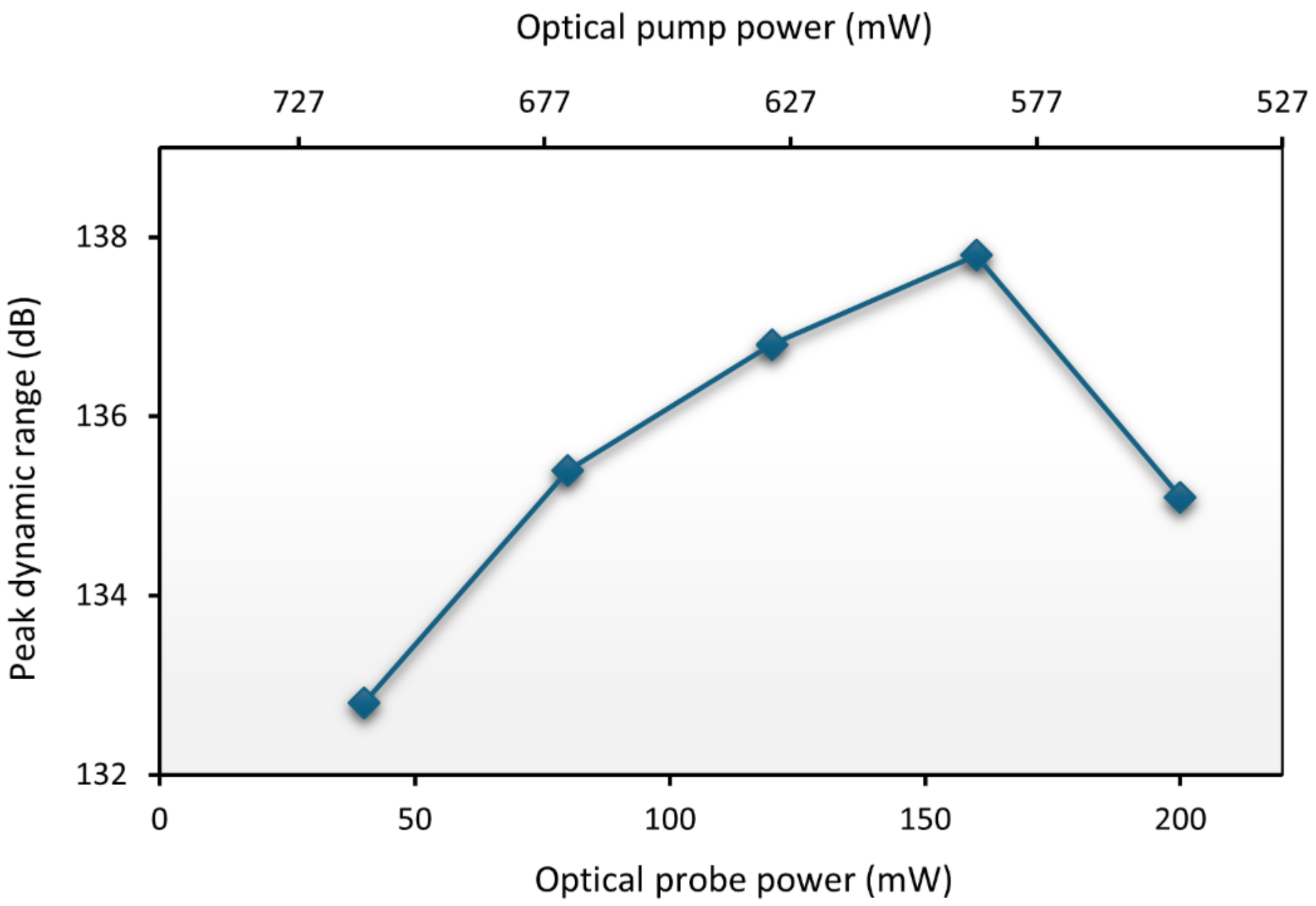

**Figure S2.** Peak dynamic range as a function of the optical pump and probe power levels driving the terahertz source and detector, respectively. The results are obtained for a THz-TDS scan range of 200 ps and an acquisition time of 60 s. In each case, the total optical power budget is fixed at 900 mW. While increasing the optical pump power enhances the generated and the detected signal amplitudes, the increased noise associated with higher optical probe powers reduces the dynamic range.